\documentclass[conference]{IEEEtran}
\usepackage{cite}
\usepackage{amsmath,amssymb,amsfonts}
\usepackage{amsthm}
\usepackage[ruled]{algorithm2e}
\usepackage{algpseudocode}
\usepackage{graphicx}
\usepackage{textcomp}
\usepackage{xcolor}
\usepackage{subcaption}
\usepackage{multirow}
\usepackage{booktabs}
\usepackage{makecell}
\usepackage{bbm}

\newcommand{\argmax}{\mathop{\mathrm{argmax}}}
\newcommand{\linebreakand}{%
  \end{@IEEEauthorhalign}
  \hfill\mbox{}\par
  \mbox{}\hfill\begin{@IEEEauthorhalign}
}

\begin{document}
\title{Discovering Command and Control (C2) Channels on Tor and Public Networks Using Reinforcement Learning}

\author{
        Cheng Wang$^{a}$$^{*}$,
        Christopher Redino$^{a}$,
        Abdul Rahman$^{a}$,
        Ryan Clark$^{a}$,\\
        Daniel Radke$^{a}$,
        Tyler Cody$^{b}$,
        Dhruv Nandakumar$^{a}$,
        Edward Bowen$^{a}$\\
        \small $^{a}$Deloitte \& Touche LLP \\
        \small $^{b}$National Security Institute, Virginia Tech \\
        \small $^{*}$Corresponding author: chengwang@deloitte.com \\
}


\maketitle

\begin{abstract}
Command and control (C2) channels are an essential component of many types of cyber attacks, as they enable attackers to remotely control their malware-infected machines and execute harmful actions, such as propagating malicious code across networks, exfiltrating confidential data, or initiating distributed denial of service (DDoS) attacks. Identifying these C2 channels is therefore crucial in helping to mitigate and prevent cyber attacks. However, identifying C2 channels typically involves a manual process, requiring deep knowledge and expertise in cyber operations. In this paper, we propose a reinforcement learning (RL) based approach to automatically emulate C2 attack campaigns using both the normal (public) and the Tor networks. In addition, payload size and network firewalls are configured to simulate real-world attack scenarios. Results on a typical network configuration show that the RL agent can automatically discover resilient C2 attack paths utilizing both Tor-based and conventional communication channels, while also bypassing network firewalls.

\end{abstract}


\section{Introduction}
A command and control (C2) channel is the communication channel that attackers use to remotely control their malware-infected devices, allowing them to issue commands to the devices or to receive information from them. A C2 channel enables attackers to maintain control over their compromised machines and carry out a variety of malicious activities, such as stealing data, launching ransomware attacks, and spreading malware to other systems.
Moreover, C2 is not just limited to individual attacks. It can be used by advanced persistent threat (APT) groups, which are state-sponsored or well-resourced organizations that engage in long-term, sophisticated attacks against specific entities. APT groups often use C2 to maintain persistence within a target's network and to exfiltrate sensitive information over a long period of time \cite{mitre-attack}.

Attackers often use techniques to hide their C2 communications to avoid detection. One such technique is the use of the Tor (The Onion Router) network, which is a decentralized network that enables anonymous and encrypted communication channels by routing traffic through multiple nodes. Tor is a free, open-source software allowing the easy proliferation of volunteer servers, or nodes, capable of obfuscating the origin of malicious traffic. By accessing a Tor server, an attacker can operate an unattributed C2 server and conceal their location from law enforcement or security researchers. 

This paper extends works done on applying reinforcement learning (RL) to different penetration testing tasks \cite{gangupantulu2021using, gangupantulu2021crown, cody2022discovering, huang2022exposing, 10115173}. Specifically, it follows the RL-based approach first proposed in \cite{gangupantulu2021using}, which incorporates cyber terrain into the Markov decision process (MDP). It also expands on the C2 simulation model introduced in \cite{10115173} by considering communications through both standard and Tor-based channels. To the best of our knowledge, this work is among the first studies using RL to simulate C2 communications through the Tor and the public networks. The RL-based approach presents a viable option to automate the discovery process of C2 pathways, thereby reducing the time and effort for manual examination. This can lead to more effective and  efficient decision-making on implementing appropriate security measures to prevent or mitigate cyber attacks.  
The main contributions of this study include the following:
\begin{itemize}
    \item Our paper proposes and develops a reinforcement learning model that incorporates the specific features of cyber defense terrain, such as firewalls, to automatically discover C2 attack pathways through both public and Tor-based communication channels.
    \item Our experiments demonstrate that the proposed RL approach is effective in helping identify strategic points in C2 attack paths and evading firewalls on a commonly-used network setting.
\end{itemize}

The remainder of the paper is organized as follows. First, we review related works in Section \ref{sec:related} and provide background information about reinforcement learning in Section \ref{sec:preliminaries}.  Then, we present the C2 attack simulation model and the details of its RL formulation in Section \ref{sec:methods}. Section \ref{sec:experiments} reports and discusses the experiment results. Finally, Section \ref{sec:conclusion} concludes the paper and suggests directions for future research.

\section{Related Work} \label{sec:related}
Reinforcement learning has emerged as a promising approach for addressing various aspects of cyber security \cite{nguyen2019deep}. One area of research focuses on using RL to automate the penetration testing process \cite{yousefi2018reinforcement, chowdhary2020autonomous, hu2020automated, tran2021deep, gangupantulu2021using}, where an RL agent learns to find the best attack path for a given network topology through trial-and-error. To better simulate the real-world scenarios, cyber terrain is introduced in \cite{gangupantulu2021using}, where firewalls represent cyber obstacles and affect rewards and transition probabilities from state to state. 
Using cyber terrain, Gangupantulu et al.  present CJA-RL \cite{gangupantulu2021crown}, an RL-based approach to perform crown jewel analysis that can be used to identify critical hosts within a network such as ideal entry points and choke points.
In \cite{cody2022discovering}, Cody el al. use RL to discover optimal exfiltration paths from compromised nodes within a network to some exit nodes that are connected to the public Internet. Following a similar approach, Huang et al. \cite{huang2022exposing} explore surveillance detection routes, in which the RL agent attempts to gather service information about target hosts and other areas of the network while exercising caution when navigating the defense terrain.

Command and control is a critical aspect of many types of cyber attacks, including malware, botnets, and ransomware \cite{vikelich2001architecture, erbacher2005extending, dittrich2007command, gardiner2014command}.
A number of studies have focused on C2 simulation, control, and infrastructure \cite{bernier2012metrics, carvalho2013mtc2, carvalho2015mira}. Other works have developed tools for automating C2 \cite{willett2015integrated, amro2022click}, such as OpenC2  \cite{mavroeidis2020nonproprietary}, which is a non-proprietary language. 
There is also an increasing body of research that has examined the use of the Tor network to conceal C2 activities \cite{brown2010resilient, casenove2014botnet, kang2015efficient, anagnostopoulos2017botnet}. While the Tor network provides a high degree of anonymity, it is not foolproof. Different techniques have been developed to detect or block  Tor traffic \cite{lashkari2017characterization, basyoni2020traffic, gurunarayanan2021improving, dodia2022exposing}. 

However, the research on employing RL for C2 discovery and maintenance remains limited. In the work presented in \cite{10115173}, a comprehensive C2 simulation model is proposed, characterized by three distinct stages in its lifecycle: infection, connection, and exfiltration. However, the model overlooks the variety of communication channels attackers might utilize for enhanced stealthiness and robustness. Furthermore, it operates under the assumption that once connections are established, they will remain uninterrupted, enabling continuous exfiltration activities without the risk of loss or disruption.

\section{Background} \label{sec:preliminaries}

Reinforcement learning is a sub-field of machine learning that involves training an agent to optimize its decisions by interacting with its environment \cite{sutton2018reinforcement}. The RL environment is generally described as a Markov decision process or MDP: $(\mathcal{S}, \mathcal{A}, \mathcal{P}, r, \gamma)$, where $\mathcal{S}$ denotes the state space and $\mathcal{A}$ the action space. At each step, the RL agent takes an action $a_t$ in its current state $s_t$, and transitions to a new state $s_{t+1}$ according to the transitional probability function
$\mathcal{P}: \mathcal{S} \times \mathcal{A} \times \mathcal{S} \rightarrow [0,1]$. The agent then receives a reward of $r_{t+1}=r(s_t, a_t, s_{t+1})$ based on the reward function $r: \mathcal{S} \times \mathcal{A} \times \mathcal{S} \rightarrow \mathbb{R}$. A discount factor $\gamma \in (0, 1]$ is often used to determine the present value of future rewards.
The agent selects its action according to its policy $\pi: \mathcal{S} \times \mathcal{A} \rightarrow [0, 1]$, which is a probability distribution over all possible actions. 
Through trial-and-error, the RL agent aims to find an optimal policy $\pi^*$ that maximizes the expected \emph{return} $G_t=\sum_{k=1}^\infty \gamma^k r_{t+k}$. 

RL algorithms fall into three main categories based on their learning approach: value-based, known as critic-only, where the focus is on assessing the value of different actions; policy-based, or actor-only, which concentrate on directly determining actions to take; and actor-critic methods, which combine the strengths of both by simultaneously evaluating actions (critic) and deciding on the next action (actor).

Critical-only methods \cite{watkins1989learning, mnih2015human} learn the optimal state-action value function $Q^*(s,a)$:
\begin{align}
    Q^*(s,a)& \equiv \max_\pi Q^\pi(s,a)  \nonumber \\
    &\equiv \max_\pi\mathbb{E}_\pi\big[G_t|s_t=s, a_t=a\big],
\end{align}
by recursively solving the Bellman optimality equation:
\begin{align}
    Q^* (s,a) = \mathbb{E}_{s'} \big[ r + \gamma \max_{a'} Q^* (s',a') | s, a\big].
\end{align}
An optimal policy is then derived by choosing the action that yields the largest $Q$-value:
\begin{align}
    \pi^*(s) = \argmax_{a}Q^*(s,a).
\end{align}

In contrast, policy-based methods directly learn a parameterized policy $\pi(a|s;\theta)$, bypassing the need to estimate the value function. A performance measure $J(\theta)$ (such as the expected return) is then optimized via gradient ascent. However, these methods frequently experience issues with high variance and slow rates of convergence. 

Actor-critic algorithms integrate value-based and policy-based approaches. These methods employ an actor for action selection and a critic to assess the value function. 
In order to reduce the variance in the estimation of the policy gradient $\nabla J(\theta)$, the value function $V_\pi(s)\equiv \mathbb{E}_\pi[G_t|s_t=s]$ is frequently used as a baseline in estimating the policy gradient \cite{nguyen2019deep}. By deducting the baseline from the action-value function, we obtain an estimate of the \emph{advantage} function $A_t=Q(s_t, a_t) - V(s_t)$, representing the difference between the expected returns from a specific action and the general returns from the current state. 
The policy gradient can subsequently be estimated as 
\begin{align}
    \nabla J(\theta)  \approx \mathbb{E}\big[\nabla_{\theta}\log \pi(a_t|s_t; \theta)A_t\big].
\end{align} 

Despite the variance reduction, policy gradient methods can still experience large policy updates. These substantial updates can result in performance collapse, a situation that is especially difficult to remedy. The challenge arises because the agent might persist in gathering samples that further reinforce the ineffective policy, making recovery more difficult. 
One approach to mitigating this issue is to use off-policy methods \cite{munos2016safe, fujimoto2019off}, in which an agent learns from samples generated by a different policy than the one being updated. 

Another approach is to use a trust region optimization method \cite{schulman2015trust}, such as the  Proximal Policy Optimization (PPO) algorithm \cite{schulman2017proximal}. In particular, PPO uses a clipped surrogate objective function:
\begin{align}
    \mathcal{L}(\theta) = \mathbb{E} \Big[ \min\big[\eta_t(\theta) A_t, \mathrm{clip}\big( \eta_t(\theta), 1-\epsilon, 1+ \epsilon \big) A_t\big]\Big], \label{eq:ppo_obj}
\end{align}
where $\eta_t(\theta) = \pi_\theta(a_t|s_t) / \pi_{\theta_{\mathrm{old}}} (a_t|s_t)$ represents probability ratio of the new policy to the old policy. The advantage function $A_t$ can be approximated using the generalized advantage estimation \cite{schulman2015high},
\begin{align}
    & \hat{A}_t = \zeta_t + (\gamma\lambda) \zeta_{t+1}+\cdots + (\gamma\lambda)^{T-t+1}\zeta_{T-1},\\
    & \mathrm{where\;} \zeta_t = r_t + \gamma V(s_{t+1}) - V(s_t).
\end{align}
In practice, an entropy bonus $\beta H(\theta)$ is frequently incorporated into the objective function \eqref{eq:ppo_obj} to promote exploration, where $\beta$ serves as a coefficient. 


\section{Methods} \label{sec:methods}

In this section, we provide an in-depth explanation of the reinforcement learning model for identifying C2 channels.
It is worth noting that while the development of our model necessitates certain assumptions that may not precisely mirror reality, it is underpinned by a data-driven approach. The inputs to the model, derived from scan data (such as Nmap or Nessus), offer flexibility for future refinements to enhance its alignment with real-world conditions.

\subsection{Attack Simulation Overview}
Building upon \cite{10115173}, we model a command and control operation as a multi-stage process encompassing (i) an initial infection stage, (ii) a subsequent establishment of a connection, and (iii) a final stage of data exfiltration. Initially, the attacker aims to establish a presence on the targeted system by leveraging existing vulnerabilities. Subsequently, the attacker attempts to set up communications with the C2 server to receive additional instructions, such as encrypting or uploading specific files. Connections to C2 can be made through the public internet or the Tor network. Upon discovering critical data on the target system, the attacker initiates the transfer of data packets from the compromised hosts to the remote C2 server through the previously established communication link. 
An attack is categorized as successful when it manages to complete all three stages within a specified timeframe.

\subsubsection{Infection Stage}
In the initial infection stage, the agent has the option to either conduct a \emph{subnet scan} or execute an \emph{exploit} action on a chosen target. Performing a subnet scan uncovers not only the devices within the same subnet but also identifies devices with specific services in nearby subnets.
Each exploit action corresponds to a known Common Vulnerabilities and Exposures (CVE) vulnerability, with its success contingent on the existence of a particular service or process and the type of operating system on the target host. A prerequisite for exploiting a host is its prior discovery via a subnet scan. Similarly, performing a subnet scan from a certain host requires first gaining access to that host. Consequently, infiltrating a high-value target typically involves the discovery and subsequent exploitation of several intermediary hosts.

\subsubsection{Connection Stage}
After successfully compromising a sensitive host, the agent might proceed to establish connections by initiating a \emph{connect} action. This involves sending a small packet to the C2 server over the public Internet or the Tor network. 
A connection attempt may fail with certain probability. In particular, it is assumed that connecting through the Tor network has a lower success probability than using the public web, as it requires additional configuration and can be more difficult to set up. 

For a successful connection to the C2 server, the outbound traffic must traverse each firewall from the host's subnet to the Internet. Furthermore, the connection will be blocked if any of these firewalls have been updated with security patches since the host was first infected.

Firewalls undergo periodic updates or are updated in response to the detection of unusual traffic patterns. These updates are scheduled based on wall-clock time, measured in seconds. This differs from the concept of time steps in Markov Decision Processes (MDPs), where each step consistently increases by one. In contrast, the increment of wall-clock time varies according to the specific action taken.  

If an excessive number of connection attempts occur, an alert will be triggered, leading to an immediate emergency update of the firewall. Additionally, if these are made using Tor then future Tor connections from the underlying host will be blocked. However the host may still be able to connect to the C2 server through the public network as its identity remains unknown. On the other hand, for non-Tor-based connections, the firewall will be updated to blacklist the IP address of C2, rendering the compromised hosts in the same subnet incapable of establishing a connection with the C2 server.

\subsubsection{Exfiltration Stage}
Once a communication channel with the C2 server is established, a specific target payload will be identified, and it can be uploaded in parts, one at a time, using either the established public channel or a Tor-based channel. The attack is complete when the entire payload has been successfully uploaded.

Similar to the connecting stage, outbound traffic is inspected by firewalls. 
Alerts will be triggered if the egress volume or the active uploading time is too high during a given monitoring window (e.g., five minutes). Depending on the connection mode, this can lead to either the blocking of Tor or the isolation of the initiating host. 

Therefore, to successfully send out the entire payload, the agent needs to exercise caution when uploading. In particular, it may need to pause the uploading process by taking other actions such as \emph{sleep}. However, if the idle time is too long, then the established connection will automatically drop and the agent will have to reconnect to the C2 server before it can upload again.

\subsection{Reinforcement Learning Formulation}\label{sec:rl}
In this section we specify the essential elements of the MDP environment for the RL agent. 
\subsubsection{States}
The state comprises the following attributes for every host: 
\begin{itemize}
    \item Subnet and local address, 
    \item Operating system (OS),
    \item Running services and processes,
    \item Whether it's discovered and the value of discovery, 
    \item Whether it's compromised and the value of infection. 
\end{itemize}
For each host, its subnet, local address and OS are one-hot encoded. Services and processes are represented by binary vectors, where one indicates that the service or process is active on the host. The discovery and infection statuses are updated from zero to one once the host is discovered and compromised, respectively. Furthermore, the discovery and infection values serve as rewards awarded to the agent following the discovery and exploitation of a host, respectively. 

Apart from the previously mentioned features, each sensitive host also includes the following:
\begin{itemize}
    \item Time since compromise,
    \item Normal connection status,
    \item Tor-based connection status,
    \item Total normal connection attempts,
    \item Total Tor-based connection attempts,
    \item Tor accessibility, 
    \item The size of the remaining payload,
    \item Total upload duration and upload data volume.
\end{itemize}
A sensitive host can be in one of three states: not connected, connected, or isolated. The measurement of its time since infection is in seconds rather than time steps. Exfiltration of data is considered complete when the remaining payload size reaches zero.

\subsubsection{Actions}
The action space encompasses  five distinct types of actions: \emph{subnet scan} and \emph{exploit} actions in stage one of the attack campaign, \emph{connect} actions in stage two, \emph{upload} actions in stage three, and a \emph{sleep} action that is available throughout the campaign.

Except for the sleep action, which essentially involves doing nothing for a predefined duration, every other action is associated with a target host. A subnet scan can be performed at every compromised host, and multiple exploits, each corresponding to a specific vulnerability, can be carried out against a discovered target.

Two connecting and uploading actions are available for each sensitive host - one via the public network and the other through the Tor network. 
As there is additional overhead associated with setting up Tor (e.g., installing the Tor client), connecting through the public Internet is faster and is more likely to be successful than connecting through the Tor network. 
As for uploading, Tor is slower than the public network due to the additional encryption and routing processes involved. 
On the other hand, Tor traffic is more difficult to monitor and therefore makes exfiltration less likely to be detected by firewalls.

Table \ref{tab:actions} shows the wall time for each type of actions employed in the three-stage attack model. The simulation clock will only increase by one second for misaligned actions, such as attempting a subnet scan from a host that hasn't been compromised or uploading from a host that hasn't been connected.

\begin{table}[b]
    \centering
    \caption{Actions for the RL agent.}
    \begin{tabular}{|c|c|c|} \hline
       \textbf{Action Type} & \textbf{Stage} & \textbf{Wall time (sec)} \\ \hline
       Subnet Scan & I & 30\\
       Exploit & I & 10 \\
       Connect (standard) & II & 10 \\
       Connect (via Tor) & II & 30 \\
       Upload & III & 10\\
       Sleep & I, II, III & 60\\ \hline
    \end{tabular}
    \label{tab:actions}
\end{table}

\subsubsection{Rewards}

The design of the reward function follows the approach proposed in \cite{gangupantulu2021using}, which defines the transition probabilities and rewards of the MDP using the Common Vulnerability Scoring System (CVSS-MDP). 
In addition, as in \cite{cody2022discovering}, services running on each host are categorized into three groups, each associated with a distinct level of penalty to signify varying degrees of defense coverage. As a result, the cost of an action is determined by its type as well as the target host's services.

The agent earns positive rewards for advancing towards the goal. These include: discovering a sensitive host' subnet, exploiting the host,  connecting the host to the remote C2 server, and uploading partial payload (Table \ref{tab:rewards}). A large bonus reward of 10000 is given once the entire payload is uploaded. In the event that exfiltration activity is detected by the firewalls, the agent receives a negative reward of -1000 or -2000 for Tor-based or public network uploads, respectively.

\begin{table}[t]
    \centering
    \caption{Types of rewards.}
    \begin{tabular}{|c|r|} \hline
        \textbf{Reward Type} & \textbf{Value} \\ \hline
        Discovery &  1000\\ 
        Exploit & 1000\\ 
        Connection & 1000\\ 
        Partial Upload (per unit) & 0.1 \\ 
        Upload completion bonus & 10000 \\ \hline
    \end{tabular}
    \label{tab:rewards}
\end{table}

\begin{figure*}[t]
    \centering
    \includegraphics[width=0.98\textwidth]{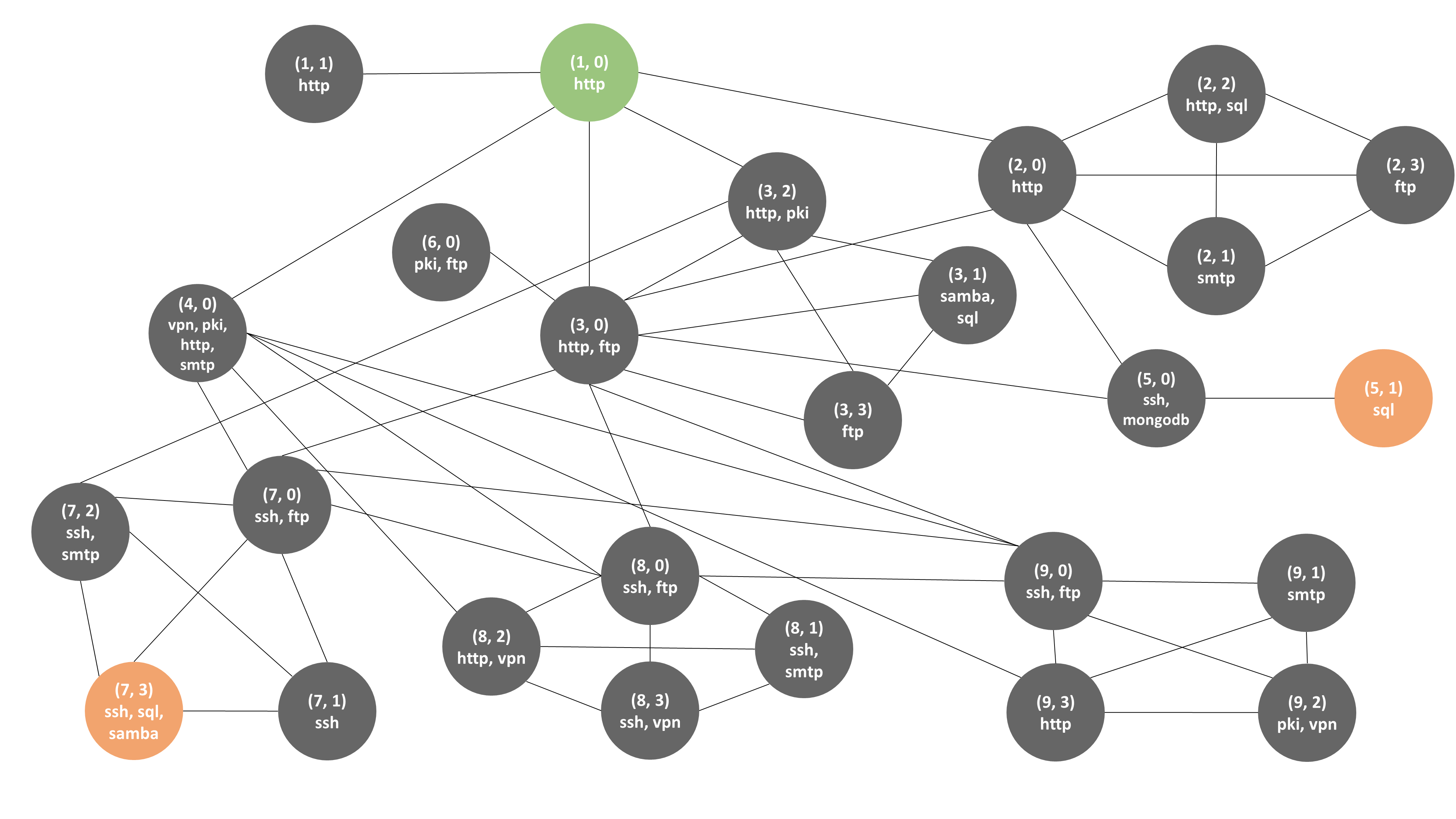}
    \caption{Network diagram with nodes IDs and services listed. Services include file transfer protocol(ftp), hypertext transfer protocol(http), virtual private network(vpn),  standard query language(sql), secure shell(ssh), Samba(samba), public key infrastructure(pki), simple mail transfer protocol(smtp) and MongoDB(mongodb). An initial foothold is gained on host (1,0) (the green node) from subnet 1. Two targets are identified and highlighted in orange.}
    \label{fig:network}
\end{figure*}

\section{Experiments} \label{sec:experiments}
In this section we introduce the experiment network, followed by a detailed account of the training procedure for the RL agent. Next, We present the training results and summarize the attack paths learned by the agent. 

\subsection{Network Description}
The network topology employed for the experiments is a commonly-used configuration in organizations. It consists of a public-facing subnet (subnet 1) and eight private subnets. Fig. \ref{fig:network} illustrates the network topology and service information for each host. The RL agent has established an initial foothold on host (1, 0), the green node in Fig. \ref{fig:network}, and aims to attack machine (5, 1) which runs standard query language (SQL) services and machine (7, 3) which runs secure shell (SSH), SQL, and Samba. Note that firewalls are placed between adjacent subnets and the Internet, even though they are not shown on the diagram. To reach the C2 server, traffic must pass every firewall along the path from the initiating host's subnet to the Internet.

\subsection{Training Details}
The RL agent undergoes episodic training using the PPO algorithm \cite{schulman2017proximal} described in section \ref{sec:preliminaries}. 
Both the actor and the critic are implemented as feed-forward neural networks, containing 128 neurons in the first layer and 64 neurons in the second. Table \ref{tab:hyperparams} lists the values of hyperparameters in the PPO algorithm.

\begin{table}[t]
    \centering
    \caption{Hyperparameters of the PPO algorithm.}
    \begin{tabular}{|l|l|} \hline
        \textbf{Hyperparameter} & \textbf{Value} \\ \hline
        Critic learning rate ($\alpha_w$) & $3\times10^{-4}$ \\ 
        Actor earning rate ($\alpha_\theta$) & $3\times10^{-5}$\\
        Discount factor ($\gamma$) & 0.99 \\
        Horizon (T) & 4096 \\ 
        Minibatch size & 64 \\
        Epochs & 5 \\
        GAE parameter ($\lambda$) & 0.95 \\
        Clipping parameter ($\epsilon$) & 0.2 \\
        Entropy coefficient ($\beta$) & 0.001 \\ \hline
    \end{tabular}
    \label{tab:hyperparams}
\end{table}

An episode ends when the the entire payload is successfully sent to the C2 server or when the host is isolated by firewalls.  
For both targets (5, 1) and (7, 3), the payload size is set to 10000 MB.

Up to three attempts are allowed to establish Tor-based and public network communication channels to C2, respectively. 
If the limit is exceeded, an alert is immediately triggered. Alerts will also be raised if the total upload volume exceeds 5000 MB or the cumulative uploading time surpasses one minute during a five-minute period. 

The success probabilities of passing through a firewall for public network and Tor-based connecting attempts are set to 0.9 and 0.75, respectively. Connection is automatically dropped after five minutes idle time and is cut off with a probability of 0.3 or 0.1 during the public network or Tor-based uploading process, respectively.

\subsection{Results}
\begin{figure}[t]
    \centering
    \includegraphics[width=0.48\textwidth]{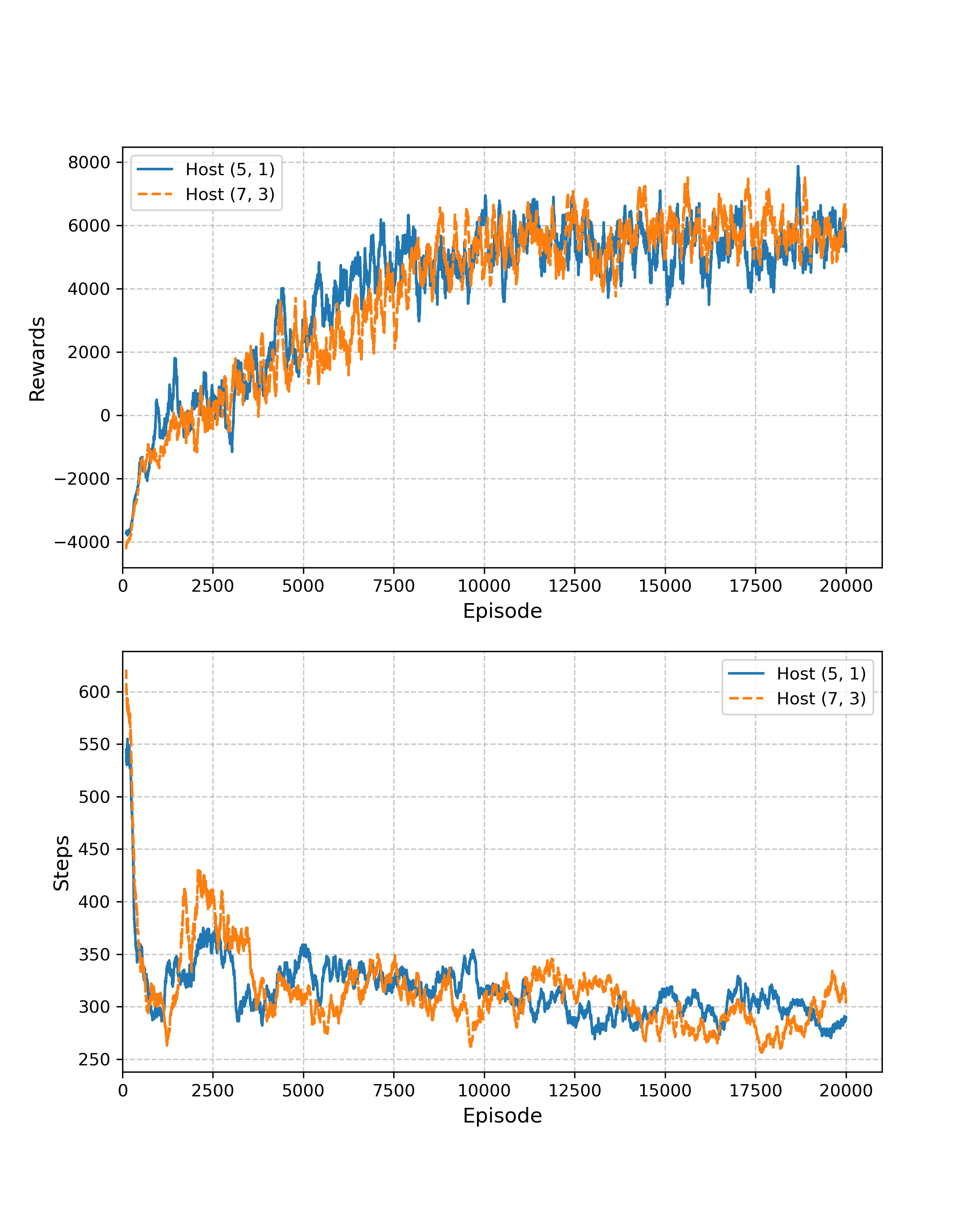}
    \caption{Average of episode rewards (top) and steps (bottoms) during the training process for targets (5,1) and (7,3), respectively.}
    \label{fig:train}
\end{figure}

The moving average of total rewards and steps of last 100 episode during training are plotted in Fig. \ref{fig:train}. It can be seen that for both targets the RL policy has converged in 20000 episodes. As training progresses, performance improves as the total rewards gradually increase to 6000 while the episode length reduces to less than 300. Given the reward structure defined in Section \ref{sec:rl} (in particular, the large reward is only given after all payload is uploaded), this indicates that the RL agent can successfully execute the attack approximately 60\% of time. 

To evaluate the final policy, 100 attack paths are generated for each target host. In both cases, 69 of those successfully sent the entire payload to the C2 server without being detected by firewalls. Table \ref{tab:summary-episodes} reports summary statistics on these trajectories. On average, 2.6 connection attempts are made via Tor and only 1.3 normal connections are made through the public channel.  

\begin{table}[b]
    \centering
    \caption{Summary statistics of successful attack paths.}
    \begin{tabular}{|c|c|c|c|c|}\hline
        ~ &  Rewards & Steps & Connections & Tor Connections\\ \hline
        Mean & 10068.3 & 296.8 & 1.3 & 2.6 \\
        Median & 10098.5 & 285.5 & 1.0 & 3.0\\ 
        Std & 308.0 & 58.9 & 1.1 & 1.1\\\hline
    \end{tabular}
    \label{tab:summary-episodes}
\end{table}

Because the learned policy operates in a random manner, the reinforcement learning agent might engage in certain actions that are not necessary. These may include scanning or exploiting less relevant hosts in other areas of the network. After examining the trajectories generated by the RL policy, key steps are identified in the attack paths to hosts (5,1) and (7,3) and are shown in Table \ref{tab:steps}. As we can see, both paths involve discovering and exploiting certain intermediate nodes in order to establish a foothold on the target system. Specifically, hosts (3, 2) and (5, 0) are also exploited on the attack path to (5, 1) and hosts (4, 0) and (7, 2) are compromised on the path to (7, 3). Once the targets are infected, multiple communication sessions are often needed to exfiltrate data to the C2 server. Each session starts with a successful connection action and can have multiple uploads. 

\begin{table}[t]
    \centering
    \caption{Key steps in attack paths to hosts (5,1) and (7,3).}
    \begin{tabular}{|l|c|l|c|} \hline
    \multicolumn{2}{|c|}{Host (5, 1)} & \multicolumn{2}{|c|}{Host (7, 3)}\\ \hline
      \textbf{Action}  & \textbf{Target} & \textbf{Action}  & \textbf{Target} \\ \hline
         Subnet Scan &  (1, 0) & Subnet Scan &  (1, 0) \\
         Exploit & (3, 2)  & Exploit & (4, 0)\\
         Subnet Scan & (3, 2) &  Subnet Scan & (4, 0) \\ 
         Exploit & (5, 0) & Exploit & (7, 2)\\ 
         Subnet Scan & (5, 0) &  Subnet Scan & (7, 2) \\
         Exploit & (5, 1) &  Exploit & (7, 3)\\
         Connect (Tor) & (5, 1) &  Connect (Tor) & (7, 3)\\
         Upload (Tor) & (5, 1) & Upload (Tor) & (7, 3) \\ 
         Connect & (5, 1) & Connect & (7, 3)\\
         Upload  & (5, 1) & Upload  & (7, 3) \\ 
     \hline
    \end{tabular}
    \label{tab:steps}
\end{table}

\begin{figure}[t]
    \centering
    \includegraphics[width=0.48\textwidth]{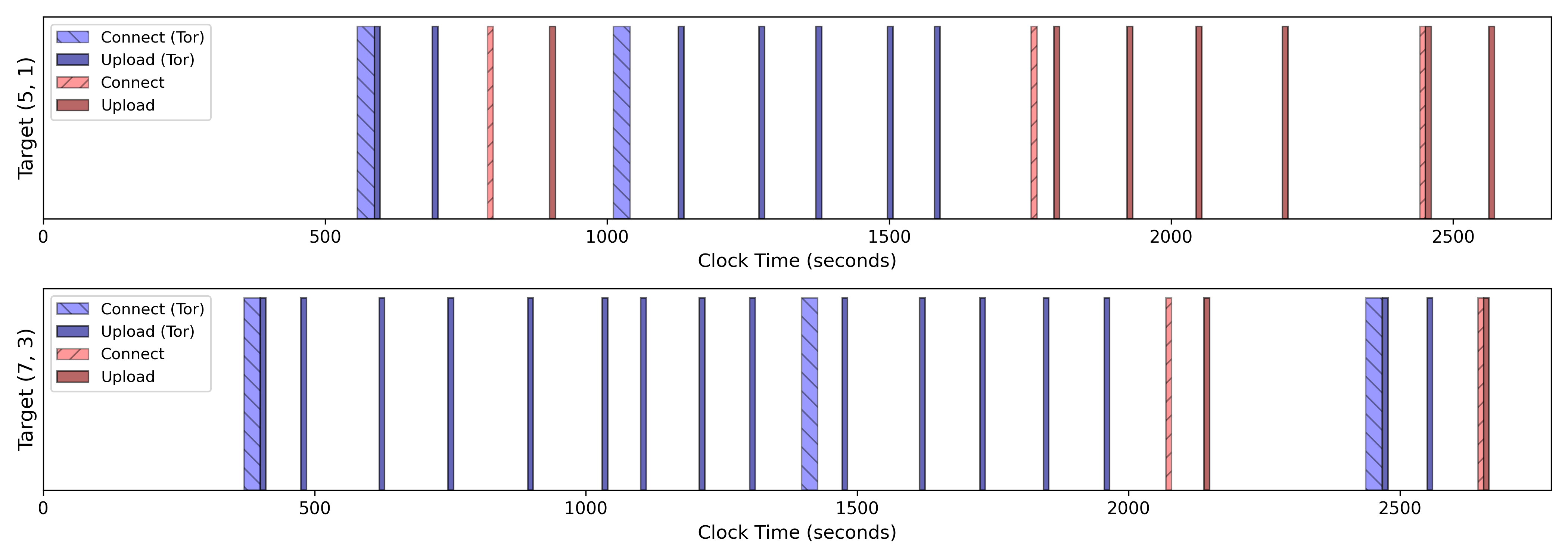}
    \caption{Timelines of connect and upload actions taken in the attack paths to host (5, 1) (top) and (7,2) (bottom).}
    \label{fig:timeline}
\end{figure}

Example timelines of communications from both infected targets to the C2 server are illustrated in Fig. \ref{fig:timeline}. As can be seen, both Tor and the public network are utilized to exfiltrate the payload. 
In addition, for both targets, the RL agent initially chooses to use Tor to communicate with the C2 server due to the added anonymity, even though it is slower as compared to the public network. Interestingly, the agent does not wait to use the public channel until the limit of Tor connections is reached or Tor is blocked. 
It is worth noting that there are regular breaks between upload actions. This is necessary to circumvent the firewalls, which will trigger alerts if high traffic volume or active time is detected. The observed traffic patterns provide additional insights into the behavior of potential attackers. This information could then be used by security analysts and operators to refine alert criteria and thereby develop more effective defensive strategies.

\section{Conclusion} \label{sec:conclusion}
In this paper, we have proposed a three-stage C2 attack model that involves both standard and Tor-based communications and have developed the corresponding reinforcement learning model. We demonstrated that the RL agent can effectively discover optimal attack paths in a given network while evading firewalls. The identified pivot points and communication patterns by the RL agent could then inform the planning and implementation of both offensive and defensive measures for security experts.

Future work may explore protocol-specific communication channels, such as use of hypertext transfer protocol(HTTP), simple mail transfer protocol(SMTP), and domain name system (DNS), as these are commonly used C2 protocols.  Another direction is to consider more sophisticated intrusion detection/prevention systems. In particular, RL could also be used as a defensive tool to refine firewall rules so that C2 activities can be detected and terminated at an early stage.

\bibliographystyle{IEEEtran}
\bibliography{ref}

\begin{thebibliography}{10}
\providecommand{\url}[1]{#1}
\csname url@samestyle\endcsname
\providecommand{\newblock}{\relax}
\providecommand{\bibinfo}[2]{#2}
\providecommand{\BIBentrySTDinterwordspacing}{\spaceskip=0pt\relax}
\providecommand{\BIBentryALTinterwordstretchfactor}{4}
\providecommand{\BIBentryALTinterwordspacing}{\spaceskip=\fontdimen2\font plus
\BIBentryALTinterwordstretchfactor\fontdimen3\font minus
  \fontdimen4\font\relax}
\providecommand{\BIBforeignlanguage}[2]{{%
\expandafter\ifx\csname l@#1\endcsname\relax
\typeout{** WARNING: IEEEtran.bst: No hyphenation pattern has been}%
\typeout{** loaded for the language `#1'. Using the pattern for}%
\typeout{** the default language instead.}%
\else
\language=\csname l@#1\endcsname
\fi
#2}}
\providecommand{\BIBdecl}{\relax}
\BIBdecl

\bibitem{mitre-attack}
\BIBentryALTinterwordspacing
(2021) Mitre att\&ck framework\textregistered. [Online]. Available:
  \url{https://attack.mitre.org}
\BIBentrySTDinterwordspacing

\bibitem{gangupantulu2021using}
R.~Gangupantulu, T.~Cody, P.~Park, A.~Rahman, L.~Eisenbeiser, D.~Radke, and
  R.~Clark, ``Using cyber terrain in reinforcement learning for penetration
  testing,'' \emph{Submitted ACM ASIACCS 2022}, 2021.

\bibitem{gangupantulu2021crown}
R.~Gangupantulu, T.~Cody, A.~Rahman, C.~Redino, R.~Clark, and P.~Park, ``Crown
  jewels analysis using reinforcement learning with attack graphs,''
  \emph{arXiv preprint arXiv:2108.09358}, 2021.

\bibitem{cody2022discovering}
T.~Cody, A.~Rahman, C.~Redino, L.~Huang, R.~Clark, A.~Kakkar, D.~Kushwaha,
  P.~Park, P.~Beling, and E.~Bowen, ``Discovering exfiltration paths using
  reinforcement learning with attack graphs,'' \emph{arXiv preprint
  arXiv:2201.12416}, 2022.

\bibitem{huang2022exposing}
L.~Huang, T.~Cody, C.~Redino, A.~Rahman, A.~Kakkar, D.~Kushwaha, C.~Wang,
  R.~Clark, D.~Radke, P.~Beling \emph{et~al.}, ``Exposing surveillance
  detection routes via reinforcement learning, attack graphs, and cyber
  terrain,'' \emph{arXiv preprint arXiv:2211.03027}, 2022.

\bibitem{10115173}
C.~Wang, A.~Kakkar, C.~Redino, A.~Rahman, A.~S, R.~Clark, D.~Radke, T.~Cody,
  L.~Huang, and E.~Bowen, ``Discovering command and control channels using
  reinforcement learning,'' in \emph{SoutheastCon 2023}, 2023, pp. 685--692.

\bibitem{nguyen2019deep}
T.~T. Nguyen and V.~J. Reddi, ``Deep reinforcement learning for cyber
  security,'' \emph{arXiv preprint arXiv:1906.05799}, 2019.

\bibitem{yousefi2018reinforcement}
M.~Yousefi, N.~Mtetwa, Y.~Zhang, and H.~Tianfield, ``A reinforcement learning
  approach for attack graph analysis,'' in \emph{2018 17th IEEE International
  Conference On Trust, Security And Privacy In Computing And
  Communications/12th IEEE International Conference On Big Data Science And
  Engineering (TrustCom/BigDataSE)}.\hskip 1em plus 0.5em minus 0.4em\relax
  IEEE, 2018, pp. 212--217.

\bibitem{chowdhary2020autonomous}
A.~Chowdhary, D.~Huang, J.~S. Mahendran, D.~Romo, Y.~Deng, and A.~Sabur,
  ``Autonomous security analysis and penetration testing,'' in \emph{2020 16th
  International Conference on Mobility, Sensing and Networking (MSN)}.\hskip
  1em plus 0.5em minus 0.4em\relax IEEE, 2020, pp. 508--515.

\bibitem{hu2020automated}
Z.~Hu, R.~Beuran, and Y.~Tan, ``Automated penetration testing using deep
  reinforcement learning,'' in \emph{2020 IEEE European Symposium on Security
  and Privacy Workshops (EuroS\&PW)}.\hskip 1em plus 0.5em minus 0.4em\relax
  IEEE, 2020, pp. 2--10.

\bibitem{tran2021deep}
K.~Tran, A.~Akella, M.~Standen, J.~Kim, D.~Bowman, T.~Richer, and C.-T. Lin,
  ``Deep hierarchical reinforcement agents for automated penetration testing,''
  \emph{arXiv preprint arXiv:2109.06449}, 2021.

\bibitem{vikelich2001architecture}
D.~Vikelich, D.~Levin, and J.~Lowry, ``Architecture for cyber command and
  control: experiences and future directions,'' in \emph{Proceedings DARPA
  Information Survivability Conference and Exposition II. DISCEX'01},
  vol.~1.\hskip 1em plus 0.5em minus 0.4em\relax IEEE, 2001, pp. 155--164.

\bibitem{erbacher2005extending}
R.~F. Erbacher, ``Extending command and control infrastructures to cyber
  warfare assets,'' in \emph{2005 IEEE International Conference on Systems, Man
  and Cybernetics}, vol.~4.\hskip 1em plus 0.5em minus 0.4em\relax IEEE, 2005,
  pp. 3331--3337.

\bibitem{dittrich2007command}
D.~Dittrich and S.~Dietrich, ``Command and control structures in malware,''
  \emph{Usenix magazine}, vol.~32, no.~6, 2007.

\bibitem{gardiner2014command}
J.~Gardiner, M.~Cova, and S.~Nagaraja, ``Command \& control: Understanding,
  denying and detecting-a review of malware c2 techniques, detection and
  defences,'' \emph{arXiv preprint arXiv:1408.1136}, 2014.

\bibitem{bernier2012metrics}
M.~Bernier, S.~Leblanc, B.~Morton, E.~Filiol, and R.~Erra, ``Metrics framework
  of cyber operations on command and control,'' in \emph{Proceedings of the
  11th European Conference on Information Warfare and Security}.\hskip 1em plus
  0.5em minus 0.4em\relax Laval, France. Academic Publishing International Ltd,
  2012, pp. 53--62.

\bibitem{carvalho2013mtc2}
M.~Carvalho, T.~C. Eskridge, L.~Bunch, A.~Dalton, R.~Hoffman, J.~M. Bradshaw,
  P.~J. Feltovich, D.~Kidwell, and T.~Shanklin, ``Mtc2: A command and control
  framework for moving target defense and cyber resilience,'' in \emph{2013 6th
  International Symposium on Resilient Control Systems (ISRCS)}.\hskip 1em plus
  0.5em minus 0.4em\relax IEEE, 2013, pp. 175--180.

\bibitem{carvalho2015mira}
M.~Carvalho, T.~C. Eskridge, K.~Ferguson-Walter, and N.~Paltzer, ``Mira: a
  support infrastructure for cyber command and control operations,'' in
  \emph{2015 Resilience Week (RWS)}.\hskip 1em plus 0.5em minus 0.4em\relax
  IEEE, 2015, pp. 1--6.

\bibitem{willett2015integrated}
K.~D. Willett, ``Integrated adaptive cyberspace defense: Secure
  orchestration,'' in \emph{Proc. Int. Command Control Res. Technol.
  Symp.(ICCRTS)}, 2015, pp. 1--13.

\bibitem{amro2022click}
A.~Amro and V.~Gkioulos, ``From click to sink: Utilizing ais for command and
  control in maritime cyber attacks,'' in \emph{European Symposium on Research
  in Computer Security}.\hskip 1em plus 0.5em minus 0.4em\relax Springer, 2022,
  pp. 535--553.

\bibitem{mavroeidis2020nonproprietary}
V.~Mavroeidis and J.~Brule, ``A nonproprietary language for the command and
  control of cyber defenses--openc2,'' \emph{Computers \& Security}, vol.~97,
  p. 101999, 2020.

\bibitem{brown2010resilient}
D.~Brown, ``Resilient botnet command and control with tor,'' \emph{DEF CON},
  vol.~18, p. 105, 2010.

\bibitem{casenove2014botnet}
M.~Casenove and A.~Miraglia, ``Botnet over tor: The illusion of hiding,'' in
  \emph{2014 6th International Conference On Cyber Conflict (CyCon
  2014)}.\hskip 1em plus 0.5em minus 0.4em\relax IEEE, 2014, pp. 273--282.

\bibitem{kang2015efficient}
L.~Kang, ``Efficient botnet herding within the tor network,'' \emph{Journal of
  Computer Virology and Hacking Techniques}, vol.~11, pp. 19--26, 2015.

\bibitem{anagnostopoulos2017botnet}
M.~Anagnostopoulos, G.~Kambourakis, P.~Drakatos, M.~Karavolos, S.~Kotsilitis,
  and D.~K. Yau, ``Botnet command and control architectures revisited: Tor
  hidden services and fluxing,'' in \emph{Web Information Systems
  Engineering--WISE 2017: 18th International Conference, Puschino, Russia,
  October 7-11, 2017, Proceedings, Part II 18}.\hskip 1em plus 0.5em minus
  0.4em\relax Springer, 2017, pp. 517--527.

\bibitem{lashkari2017characterization}
A.~H. Lashkari, G.~Draper-Gil, M.~S.~I. Mamun, A.~A. Ghorbani \emph{et~al.},
  ``Characterization of tor traffic using time based features.'' in
  \emph{ICISSp}, 2017, pp. 253--262.

\bibitem{basyoni2020traffic}
L.~Basyoni, N.~Fetais, A.~Erbad, A.~Mohamed, and M.~Guizani, ``Traffic analysis
  attacks on tor: A survey,'' in \emph{2020 IEEE International Conference on
  Informatics, IoT, and Enabling Technologies (ICIoT)}.\hskip 1em plus 0.5em
  minus 0.4em\relax IEEE, 2020, pp. 183--188.

\bibitem{gurunarayanan2021improving}
A.~Gurunarayanan, A.~Agrawal, A.~Bhatia, and D.~K. Vishwakarma, ``Improving the
  performance of machine learning algorithms for tor detection,'' in \emph{2021
  International Conference on Information Networking (ICOIN)}.\hskip 1em plus
  0.5em minus 0.4em\relax IEEE, 2021, pp. 439--444.

\bibitem{dodia2022exposing}
P.~Dodia, M.~AlSabah, O.~Alrawi, and T.~Wang, ``Exposing the rat in the tunnel:
  Using traffic analysis for tor-based malware detection,'' in
  \emph{Proceedings of the 2022 ACM SIGSAC Conference on Computer and
  Communications Security}, 2022, pp. 875--889.

\bibitem{sutton2018reinforcement}
R.~S. Sutton and A.~G. Barto, \emph{Reinforcement learning: An
  introduction}.\hskip 1em plus 0.5em minus 0.4em\relax MIT press, 2018.

\bibitem{watkins1989learning}
C.~J. C.~H. Watkins, ``Learning from delayed rewards,'' 1989.

\bibitem{mnih2015human}
V.~Mnih, K.~Kavukcuoglu, D.~Silver, A.~A. Rusu, J.~Veness, M.~G. Bellemare,
  A.~Graves, M.~Riedmiller, A.~K. Fidjeland, G.~Ostrovski \emph{et~al.},
  ``Human-level control through deep reinforcement learning,'' \emph{Nature},
  vol. 518, no. 7540, pp. 529--533, 2015.

\bibitem{munos2016safe}
R.~Munos, T.~Stepleton, A.~Harutyunyan, and M.~Bellemare, ``Safe and efficient
  off-policy reinforcement learning,'' \emph{Advances in neural information
  processing systems}, vol.~29, 2016.

\bibitem{fujimoto2019off}
S.~Fujimoto, D.~Meger, and D.~Precup, ``Off-policy deep reinforcement learning
  without exploration,'' in \emph{International conference on machine
  learning}.\hskip 1em plus 0.5em minus 0.4em\relax PMLR, 2019, pp. 2052--2062.

\bibitem{schulman2015trust}
J.~Schulman, S.~Levine, P.~Abbeel, M.~Jordan, and P.~Moritz, ``Trust region
  policy optimization,'' in \emph{International conference on machine
  learning}.\hskip 1em plus 0.5em minus 0.4em\relax PMLR, 2015, pp. 1889--1897.

\bibitem{schulman2017proximal}
J.~Schulman, F.~Wolski, P.~Dhariwal, A.~Radford, and O.~Klimov, ``Proximal
  policy optimization algorithms,'' \emph{arXiv preprint arXiv:1707.06347},
  2017.

\bibitem{schulman2015high}
J.~Schulman, P.~Moritz, S.~Levine, M.~Jordan, and P.~Abbeel, ``High-dimensional
  continuous control using generalized advantage estimation,'' \emph{arXiv
  preprint arXiv:1506.02438}, 2015.

\end{thebibliography}

\end{document}